# A flux calibration method for remote sensing satellites using stars

XU Chun

[1]State Key Laboratory of Infrared Physics,
[2]Joint Center of SHAO and SITP for Infrared Astronomical Instrumentation,
Shanghai Institute of Technical Physics, Chinese Academy of Sciences, Shanghai 200083, China
Email: chun.xuu@gmail.com

**Abstract:** Star surveys and model analyses show that many stars have absolute stable fluxes as good as 3% in 0.3-35 μm wavebands and about 1% in the visible wavebands. The relative flux calibrations between stars are better than 0.2%. Some stars have extremely stable fluxes and can be used as long term flux calibration sources. Stellar brightness is several orders of magnitude lower than most ground objects while the stars do not usually appear in remote sensing cameras, which makes the stars inappropriate for being calibration sources. The calibration method using stars discussed in this paper is through a mini-camera attached to remote sensing satellite. The mini-camera works at similar wavebands as the remote sensing cameras and it can observe the stars and the ground objects alternatively. High signal-to-noise ratio is achieved for the relatively faint stars through longer exposure time. Simultaneous precise cross-calibration is obtained as the mini-camera and remote sensing cameras look at the ground objects at the same time. The fluxes from the stars used as calibration standards are transferred to the remote sensing cameras through this procedure. Analysis shows that a 2% accurate calibration is possible.

**Key words:** flux calibration; stellar; remote sensing satellite;
**Chinese Library Catalogue number:** P715.6        **Document Code**: A        **PACS**：07.60.Dq

## Introduction

Remote sensing satellites play essentially important roles in the research areas of geography, geology, natural resource, climate and in the application areas of agriculture, weather forecast and militaries. The quantitative applications of the remote sensing data are very active research fields in the modern space remote sensing instruments and the flux calibration is a key step among the calibrations. Flux calibration is a procedure to obtain one-to-one relationships between the remote sensing data and the target's radiation powers. Due to interferences from different physical processes and variable ambient temperatures and uncertain radiations in space, flux calibration is always one of the hardest problems in remote sensing. The best flux calibration accuracies are around 2-3% and rarely surpass 1% for space remote sensing instruments[1]. The flux calibration accuracies for national remote sensing instruments are even lower, for example, the flux calibration uncertainty for the High-frequency radiometric calibration for wide field-of-view sensor of GF-1 satellite is 5.3%[2] and that of the 3$^{rd}$ Earth resources satellite at visible light is 4.7%[3]. The laboratory flux calibration uncertainty for Oceanographic satellite HY-1A is as high as 25% so vicarious calibration is adapted as the flux calibration plan for HY-1A[4]. Some research fields such as the long time variabilities of the earth climate, the meterological environmental and geological changes due to the sun radiation variabilities, will need not only precise flux calibration sources but also long term stable calibration sources.

Two problems need to be resolved in order to reach a high flux calibration: one is to find a precise and stable calibration source and the other is to develop a proper calibration method. Man made calibration sources such as calibration lamps, black bodies, precision detectors all suffer certain performance degradations as used in space.

Natural calibration sources such as the ground calibration fields, the Sun, the Moon whose fluxes are not very stable and these objects are only observable at certain satellite orbits. The Sun is often chosen to be a calibration source in visible wavelength through a diffuse reflection board but the uncertainty of the solar flux is around 3%[5] and the diffuse reflection board suffers orbital performance degradations. The Moon is sometimes used as a calibration source while it's pre-determined instantaneous flux values are uncertain to 10%[6] due to the Sun's radiation angle, moon phase and the satellite's orbital positions. Guo Qian et al.[7,8] calibrated the FY-2 meterological satellite data using the Moon and reached 1-3.5K (approximately 5-17%) uncertainties in infrared wavelengths. The flux calibration uncertainties of the ground calibration fields[2,9] are from the uncertainties of the ground flux measurements and the radiation absorptions through the atmosphere. The calibration uncertainties from vicarious calibration[4] through other satellites are from two parts: one is the flux calibration precision of the reference satellite and the other is from the observation simultaneity, orientation and wavebands consistancies of the two satellites. All the calibration methods mentioned above can not be executed at any time and the calibration procedures usually use some work time of the remote sensing instruments.

Here we propose a new flux calibration method for the remote sensing satellites using the stars. The dim but stable stellar fluxes are transferred to the ground target through this new method so the flux calibrations are performed. This new calibration will not use the work time of the remote sensing instruments. The calibration is independent of the satellite's current calibration methods so it can either be an independent calibration or a supportive calibration for the satellite. We will discuss in detail this calibration method and the calibration errors in the following chapters.

## 1  The problems and resolutions for the flux calibration using stars

The flux calibration method using stars is primarily used in astronomical satellites and astronomical observations. It is rarely used by remote sensing satellites because: 1) the stars are dimmer than the ground targets by several orders of magnitude so they don't have enough signal-to-noise ratios in the normal observation; 2) most remote sensing satellites observe the earth and it's not convient for them to look at the stars; 3) the stellar fluxes are primarily measured at the astronomically important wavebands so they do not fit most remote sensing satellites; 4) stars are point sources so their angular sizes are much smaller than the point spread functions of the remote sensing cameras while the ground targets are extended sources so there are certain technical issues to resolve to obtain the right fluxes for the two objects.

The brightness of a star is described with stellar magnitudes in astronomical catalogues. The stellar magnitude is the illuminance of the star at the observing site. A star is a point source so its illuminance is labeled as $W/cm^2$. Ground targets are usually extended objects so their brightness is labeled with $W/cm^2/srd$ which is illuminance per solid angle. The illuminance of extended objects is their brightness multiplying with their solid angles. If only the illumination to one pixel is considered, the solid angle used is just the pixel's angular resolution. Table 1 compares the star's and ground target's brightness at different pixel angular resolutions in visible (0.5 μm) and infrared (5 μm) wavebands. The ground targets appear to have similar brightness as stars only if the camera's pixel resolutions are very higher (<1"). In most cases the remote sensing camera's pixel resolutions are bigger than 5" so the ground targets appear much brighter than stars. There are problems for flux calibration using stars when stellar brightness is too small, for example, low signal-to-noise ratios from the stars will result in higher statistical errors. In our new flux calibration method we will extend the exposure time for star observations so high signal to noise images will be obtained.

Table 1 Comparison of the ground brightness per pixel with stellar brightness

| visible 0.5μm, ground reflection 36% | | infrared 5μm, 280K ground radiation | |
|---|---|---|---|
| pixel resolution | Stellar magnitude | pixel resolution | infrared AB magnitude |
| 5″ | -1.5 | 5″ | -0.8 |
| 3″ | -0.3 | 3″ | 0.3 |
| 1″ | 2.1 | 1″ | 2.7 |
| 0.5″ | 3.6 | 0.5″ | 4.2 |
| 0.3″ | 4.7 | 0.3″ | 5.3 |
| 0.1″ | 7.1 | 0.1″ | 7.7 |

Remote sensing satellites are usually oriented to the earth so flux calibrations using stars are not favored for most remote sensing satellites. Very few remote sensing satellites used stars to perform the flux calibrations. The IKONOS performed flux calibration with stars since this satellite has higher angular resolution and it is flexible to observe stars[10]. The flux calibration method we proposed here makes use of a two dimensional rotation platform and it can switch between the stars and the ground targets conveniently.

The stellar fluxes in the stellar catalogues are mostly taken at the wavebands of astronomical implications. These wavebands are not matched by most remote sensing satellites. It is necessary to obtain the stars' fine spectra in order to perform the flux calibrations for the remote sensing satellites. Such a work should rely on the spectral models the astronomers calculated because most stars don't have fine spectral observational data. Kurucz and Castelli et. al. [11,12] had calculated thousands of stellar radiative models according to the stellar population and evolution. The calculated spectra reaches 1/10000 spectral resolution and covers violet, visible and infrared wavebands so are widely used in astronomical data analysis and flux calibrations. We will retrieve star's multi-wavelength observational data and match the data with the stellar models to obtain the full spectra of the stars.

The basic principle of the new method we proposed for flux calibration of the remote sensing satellites is described below. A small camera is selected as the calibration camera and is mounted on the remote sensing satellite through a two dimensional rotating platform so the calibration camera can observe the calibration stars and the ground targets conveniently. Calibration stars are the stars with stable fluxes and are evenly distributed in the sky. Calibration camera has a proper field of view to cover a sky area with some calibration stars in so to make the star observation conveniently. Calibration camera usually has smaller field of view than the remote sensing camera but it is rotatable so it can cover the remote sensing camera's field of view with multiple observations. During flux calibration, the calibration camera observe the stars, then it observe the ground targets to transfer the stellar fluxes to the ground targets. Remote sensing camera also observes these ground targets so it can be calibrated by these targets. The calibrated remote sensing camera then observes other ground scenes with good calibrations. While the remote sensing camera is on its normal work, the calibration camera observes between the calibration stars and the ground objects alternatively. The calibration stars to observe at a certain time are determined by the satellite's orbital position and the calibration star distributions. The ground targets to observe by the calibration camera are somehow determind by the previous ground observations so after certain ground observations all the field of view of the remote sensing camera is covered by the calibration camera. This procedure is repeated over and over. The wavebands of the calibration camera are chosen to be same or very close to those of the remote sensing camera so the fluxes can be calculated properly. The exposure times of the calibration camera to the ground objects and to the stars are different so high signal-to-noise ratios are obtained for both stellar and ground targets. The exposure time to the stars are usually much longer because the stars are dimmer. The calibration camera and remote sensing camera work parallelly. Their data are analyzed independently. The calibration observation will not take up any of the remote sensing camera's work time. This calibration method is practically a vicarious calibration. Since the calibration camera and the remote sensing camera have same observation time, same orientation to the ground targets and same or similar wavebands, this vicarious calibration is much better than other vicarious calibrations between different satellites. We will analyze the calibration precision in the whole calibration process to assure the feasibility of this flux calibration method.

## 2  Analysis of all calibration errors

### 2.1  Error propagations in the calibration

According to the previous discussions, the flux calibration can be arranged into 4 steps: calibration camera observes the stars; calibration camera observes the ground target A; remote sensing camera observes ground target A simultaneously; remote sensing camera observes ground target B or any targets other than target A. The flux calibration precision is determined by the calibration precision of target B.

Assume that the stellar spectrum is $F_s(\lambda)$ near the observing wavelength while that of target A is $F_A(\lambda)$ and that of target B is $F_B(\lambda)$. Assume that the spectral response curve of the calibration camera is $R_c(\lambda)$ and the system responsibility is $\alpha_c$ while those of the remote sensing camera are $R_r(\lambda)$ and $\alpha_r$. Then we have:

$$N_c = \alpha_c \int F_s(\lambda) R_c(\lambda) d\lambda \quad (1)$$

$$N_{Ac} = \alpha_c \int F_A(\lambda) R_c(\lambda) d\lambda \quad (2)$$

$$N_{Ar} = \alpha_r \int F_A(\lambda) R_r(\lambda) d\lambda \quad (3)$$

$$N_{Br} = \alpha_r \int F_B(\lambda) R_r(\lambda) d\lambda \quad (4)$$

Here $N_c$ and $N_{Ac}$ are the outputs of the calibration camera when it observes star and ground target A, $N_{Ar}$ and $N_{Br}$ are the outputs of the remote sensing camera when it observes ground target A and B. The exposure times are not shown in these equations because they are very accurate and will not result in measurement errors. Take derivatives of eq (1) and we will have:

$$\frac{\delta N_c}{N_c} = \frac{\delta \alpha_c}{\alpha_c} + \frac{\int \delta F_s(\lambda) R_c(\lambda) d\lambda}{\int F_s(\lambda) R_c(\lambda) d\lambda} + \frac{\int F_s(\lambda) \delta R_c(\lambda) d\lambda}{\int F_s(\lambda) R_c(\lambda) d\lambda} \quad (5)$$

Here $\delta N_c/N_c$ is the difference between the calibration camera's output and the theoretical expectation, which is the total error of observing a star. The errors are from uncertainties of the calibration camera's responsibility $\delta \alpha_c$ and that of its spectral response curve $\delta R_c(\lambda)$ and that of the stellar flux $\delta F_s(\lambda)$. We also have similar equations like eq (5) from equations (2), (3), (4), which correspond to the differences of $\delta N_{Ac}/N_{Ac}$, $\delta N_{Ar}/N_{Ar}$, $\delta N_{Br}/N_{Br}$.

$\delta N_c/N_c$ and $\delta N_{Ac}/N_{Ac}$ are the differences between the calibration camera's output and expectation while these two values should be equal upon the measurement errors, so we have:

$$\frac{\delta N_{Ac}}{N_{Ac}} = \frac{\delta N_c}{N_c} + \delta m1 \qquad (6)$$

Here $\delta m1$ is the random error in measuring the stellar flux and the flux of target A, it depends on the signal-to-noise ratios so $\delta m1$ will be reduced upon multiple measurements. Combining eq (5) and (6) gives:

$$\frac{\int \delta F_A(\lambda)R_c(\lambda)d\lambda}{\int F_A(\lambda)R_c(\lambda)d\lambda} = \frac{\int \delta F_S(\lambda)R_c(\lambda)d\lambda}{\int F_S(\lambda)R_c(\lambda)d\lambda} + \left(\frac{\int F_S(\lambda)\delta R_c(\lambda)d\lambda}{\int F_S(\lambda)R_c(\lambda)d\lambda} - \frac{\int F_A(\lambda)\delta R_c(\lambda)d\lambda}{\int F_A(\lambda)R_c(\lambda)d\lambda}\right) + \delta m1 \qquad (7)$$

The left side of eq (7) is the calibrated flux error of target A by the calibration camera. The first term of the right side is the flux uncertainty from the star, the second term which is in the parenthesis is the error from the uncertainty of spectral response curve of the calibration camera and it will be reduced if the spectra of the stars and of target A are similar, the third term $\delta m1$ is from the measurements due to the signal-to-noise ratios of the calibration camera. We also notice that the term $\delta\alpha_c/\alpha_c$ in eq (5) disappeared in eq (7), which means that the change of responsibility of the calibration camera does not change the flux calibration.

With similar calculations we can get the calibration error that the remote sensing camera have in calibrating target B with target A. A similar equation like eq (7) is obtained:

$$\frac{\int \delta F_B(\lambda)R_r(\lambda)d\lambda}{\int F_B(\lambda)R_r(\lambda)d\lambda} = \frac{\int \delta F_A(\lambda)R_r(\lambda)d\lambda}{\int F_A(\lambda)R_r(\lambda)d\lambda} + \left(\frac{\int F_A(\lambda)\delta R_r(\lambda)d\lambda}{\int F_A(\lambda)R_r(\lambda)d\lambda} - \frac{\int F_B(\lambda)\delta R_r(\lambda)d\lambda}{\int F_B(\lambda)R_r(\lambda)d\lambda}\right) + \delta m2 \qquad (8)$$

The left side of eq (8) is the calibrated flux error of target B by the remote sensing camera. The first term of the right side is the flux uncertainty of target A, the second term which is in the parenthesis is the error from the uncertainty of spectral response curve of the remote sensing camera and it is highly reduced since the spectra of target A and Target B are usually similar, the third term $\delta m2$ is from the measurements due to the signal-to-noise ratios of the remote sensing camera. If the spectral response curves of the calibration camera and remote sensing camera are same or very similar, we can approximate the first term of the right side of eq (8) to the left term of eq (7), that is:

$$\frac{\int \delta F_A(\lambda)R_r(\lambda)d\lambda}{\int F_A(\lambda)R_r(\lambda)d\lambda} \simeq \frac{\int \delta F_A(\lambda)R_c(\lambda)d\lambda}{\int F_A(\lambda)R_c(\lambda)d\lambda} \qquad (9)$$

So we get the calibration error of target B by plugging eq (7) into eq (8):

$$\frac{\int \delta F_B(\lambda)R_r(\lambda)d\lambda}{\int F_B(\lambda)R_r(\lambda)d\lambda} = \frac{\int \delta F_S(\lambda)R_c(\lambda)d\lambda}{\int F_S(\lambda)R_c(\lambda)d\lambda} + \left(\frac{\int F_S(\lambda)\delta R_c(\lambda)d\lambda}{\int F_S(\lambda)R_c(\lambda)d\lambda} - \frac{\int F_A(\lambda)\delta R_c(\lambda)d\lambda}{\int F_A(\lambda)R_c(\lambda)d\lambda}\right) + \left(\frac{\int F_A(\lambda)\delta R_r(\lambda)d\lambda}{\int F_A(\lambda)R_r(\lambda)d\lambda} - \frac{\int F_B(\lambda)\delta R_r(\lambda)d\lambda}{\int F_B(\lambda)R_r(\lambda)d\lambda}\right)$$
$$+ \delta m1 + \delta m2 \qquad (10)$$

The flux calibration precision is mostly determined by the stellar flux precision, the onboard change of the the spectral response curves of the calibration camera and remote sensing camera, and the measurement errors. The 5 terms in the right side of eq (1) are independent and in fact they can be considered as the fractions of an array. So the total calibration errors are the square root of the squares of these 5 terms rather than their direct summations.

**2.2 The precision of the stellar fluxes**

We now discuss the precision and stability of stellar spectra. The stellar catalogues show that the radiations of many stars are very stable while stellar flux variabilities are less than 1%. So stars can be good flux calibration sources. The Hipparcos catalogue[13] published in 1997 contains more than 120,000 stars brighter than 12 magnitude and the internal flux precisions in visible wide spectral band (Hp band) are better than 0.2%. The stellar fluxes of each star are measured over 110 times by the Hipparcos satellite during the 4 year's mission so they have very good consistencies and stabilities[13]. The GAIA satellite launched in 2013 observed fainter stellar objects and obtained 500 million stellar data. The flux precisions are around 3-15 milli-magnitude (0.3-1.4% errors). The stellar photometry is generally measured through comparing with the standard star fluxes so the stellar fluxes in the catalogues can be considered as relative precisions. The comparison between stellar photometry and the international radiation units are obtained through measuring the standard star's fluxes such as the Vega and Sirius fluxes with the National Institute of Standards and Technology (or NIST) traceable ground instruments. The precision of this comparison is around 0.7%-2%[15,16,17]. Hence the accuracy of star's flux is around 2%. But the precision and stability of the star's flux is around 0.2%. The biggest errors for stars becoming good calibration sources are from the calibration steps to match the stellar fluxes with the NIST traceable standard sources. If this measurement error can be reduced in the future, the errors for the stars being calibration sources will be greatly reduced.

Other than the visible wavebands, the stars have strong and stable radiations at ultraviolet and infrared wavebands so they can be used as calibration sources in these wavebands too. The flux precision of stars are as good as 3% or even better than 1% at certain wavebands within 1-35 micrometer wavelength range.

We retrieve the stellar observational data at different wavebands from the virtual observatory Vizier[20]. A star usually has tens to hundreds observational data at different wavebands. Some observational data are from space telescopes such as Hubble,Spitzer,IRAS,WISE and AKARI and some data are from the ground telescopes such as 2MASS, SDSS and some other visible or infrared telescopes. The data from the ground telescopes are processed through the standard data processing procedures so the atmosphere effects are corrected and the space observational values are recovered. We also retrieved the stellar spectral models from Castelli & Kurucz[11、12]. 4300 ATLAS-9 stellar models and 7600 1993-year stellar models are retrieved. The fixed parameters of the stellar models are metal abundances, effective temperature and surface gravity and the fitting parameter is the solid angle of the star towards to observer. Since the stellar spectral models are the radiations from the stars while the stellar radiations detetced in the space have undergone radiation absorptions by the interstellar medium, the intersetllar absorptions have to calculate in model fitting. We selected the sparse interstellar medium absorption model by Cardelli, Clayton & Mathis[21] to correct the stellar spectra and the corresponding fitting parameter is a scale factor which is also obtained in the previous model fitting procedure. We fit all the observational data of a star with the thousands of spectral models and the interstellar models through least square fitting method and choose the best fitting parameters as the stellar spectra. Figure 1 is a fitting plot to the multi-spectral data of one star. According to the fitting results to part of the stars analyzed, the fitting errors (residuals) are mostly within 1-5%, some of them reach 20% at certain wavelengths. The large fitting errors are mostly from the inconststencies between different observational data at neighboring wavelengths. The inconststencies are either from the calibration errors between different instruments or from the variabilities of the stellar fluxes, or the stellar spectral models that are not suitable for certain stars. If the stellar data are fitted piece-by-piece within different wavelength ranges, the fitting residuals will reduce and the fitting errors will reach 1%. We will report the technical details of calibration star selections in other papers and release calibration star catalogues that cover the entire sky.

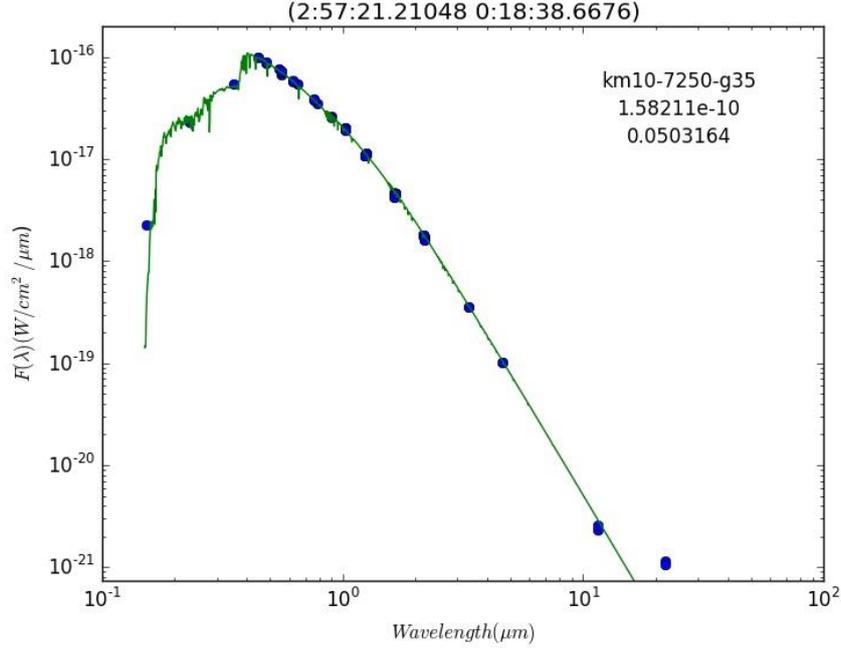

Figure 1 The fitting to one star's spectrum (Star's ICRS J2000 coordinate is 2:57:21.21 0:18:38.67). The circles are observational data at different wavebands, the thick line is theoretical model. The 3 numbers in the upper right corner of the figure are model name, effective solid angle of the star and absorption coefficient.

The flux calibration can be based on one star or several stars. It is likely that multiple stars are used as calibration sources because stars are distributed at different places in the sky while the calibration is needed at anytime. If N calibration stars are selected for the flux calibration, the brightness of ith star is $S_i \pm \delta S_i$ or it's precision is $\sigma_i = \delta S_i/S_i$ and this star is used as a calibration source for $n_i$ times in the flux calibration. If the stellar flux errors are independent, what is the flux calibration precisions for many calibration sources combined as "one" source to the precision of any single source?

The parameter the flux calibration to determine is the effective responsivity of the camera. For stars whose flux precision is σ, the best precision of the determined effective responsivity is σ too. If N calibration stars are selected and the precision of ith star is $\sigma_i$ and that star is observed $n_i$ times in calibration then the total observation time is:

$$n_t = \sum_{i=1}^{N} n_i \tag{11}$$

The total observational errors are:

$$e_t = \sqrt{\overline{(\sum_{i=1}^{N} n_i \sigma_i)(\sum_{i=1}^{N} n_i \sigma_i)}} = \sqrt{\sum_{i=1}^{N} n_i^2 \sigma_i^2} \tag{12}$$

The relationship $\overline{\sigma_i \sigma_j} = \delta_{ij}\sigma_i^2$ is used in calculation, i.e., the covariance of one star is its precision while the cross covariance between different stars are 0. Then the systematic precision of a group of N stars with $n_i$ observations each is

$$err = \frac{e_t}{n_t} = \sqrt{\sum_{i=1}^{N} n_i^2 \sigma_i^2} / \sum_{i=1}^{N} n_i \tag{13}$$

This result is consistent with the common sense. If only one calibration star i is observed so only $n_i$ is nonzero then the systematic precision is this star's precision $\sigma_i$. If each star is observed once, i.e., all $n_i$ are 1, then the systematic precision is $\sqrt{\sum_{i=1}^{N} \sigma_i^2}/N$. If each star's precision is σ, then the systematic precision is σ/$\sqrt{N}$, i.e., the precision is improved $\sqrt{N}$ times.

For a selected group of calibration stars, what is the best choice of the observation times $n_i$ so the total calibration precision is best? The best calibration precision means its derivative is 0, so

$$\delta err = \delta\left(\sqrt{\sum_{i=1}^{N} n_i^2 \sigma_i^2} / \sum_{i=1}^{N} n_i\right) = \delta\left(\sqrt{\sum_{i=1}^{N} n_i^2 \sigma_i^2} / n_t\right) = 0 \tag{14}$$

If the total observation time $n_t$ is fixed, then we have

$$\delta \sum_{i=1}^{N} n_i^2 \sigma_i^2 = \sum_{i=1}^{N} 2 n_i \delta n_i \sigma_i^2 = 0 \tag{15}$$

On the other hand, fixed $n_t$ gives

$$\delta \sum_{i=1}^{N} n_i = \sum_{i=1}^{N} \delta n_i = 0 \tag{16}$$

Equations (15) (16) show that if $n_i \sigma_i^2$ =constant then eq (15) is valid and the minimum systematic errors are achieved. In other words, the stars with better precisions are observed more times while those with poorer precisions are observed less times, as long as $n_i \sigma_i^2$ =constant, the minimum systematic errors are achieved. Put $n_i \sigma_i^2$ =constant into eq (14) and obtain the minimum total errors

$$\text{err}_{sys} = \frac{1}{\sqrt{\sum_{i=1}^{N}\left(\frac{1}{\sigma_i^2}\right)}} \tag{17}$$

If the precisions of all calibration stars are σ, then the systematic precision is σ/$\sqrt{N}$ and is consistent with eq (13).

### 2.3 The precision of calibrating the ground targets with stars

According to the previous discussions, a calibration camera that can observe the stars and ground targets can transfer the stellar fluxes to the ground targets. The signal-to-noise ratio of the calibration camera for stellar observation is given by this equation[22]

$$\text{SNR} = \frac{f_\lambda \cdot A \cdot \Delta\lambda \cdot \Delta t \cdot q_o \cdot q_e}{\sqrt{\left(f_\lambda + \Omega_{npix}(I_{sky}+I_{ins})\right) \cdot A \cdot \Delta\lambda \cdot \Delta t \cdot q_o \cdot q_e + n_{pix} \cdot I_{dark} \cdot \Delta t + n_{pix} \cdot N_{read}^2}} \tag{18}$$

Here $f_\lambda$ is stellar brightness (per wavelength), A is the effective collection area of the calibration camera, $\Delta\lambda$ is the operating waveband, $\Delta t$ is the exposour time, $q_o$ is the optical transmittance of the camera, $q_e$ is the detector's quantum efficiency, $\Omega_{npix}$ is the total solid angle towards the sky of the pixels that cover the entire star image in performing aperture photometry[23], $I_{sky}$ is the sky background brightness, $I_{ins}$ is the instrument's background radiation, $n_{pix}$ is the pixel number within the photometry aperture, $I_{dark}$ is the dark current in e⁻/s/pixel, $N_{read}$ is the read noise per pixel in electron. If the waveband is wide and $f_\lambda$ is not constant, the multiplication in the numerator and part of denominator in equation (18) can be replaced with an integral. Assume that the calibration camera's aperture diameter is 5cm, it works at 0.5 μm with 0.2 μm waveband, the detector is small focal plane array with pixel resolution of 5″, read noise is 20 electrons, dark current is 10 e⁻/s, sky background is 22 mag/square-arcsec, aperture photometry area is 16 pixels, then the calibration camera can detect 8.3 magnitude stars with signal-to-noise ratio 250 at 5 seconds exposure. This gives the measurement error of 0.4% for single exposoure. To obtain the same signal-to-noise ratio for the ground targets (approximately 0 magnitude as listed in Table 1) the exposour time is only 5 milliseconds. The other requirements for calibration camera to reach the above signal-to-noise ratio is that the precision of exposure time is better than 1 microsecond and the pointing stability is better than 5″. Both are achievable with current technologies. At longer infrared wavebands the calibration camera may need to be cooled to reduce the instrumental background radiation.

The key issue for flux calibration is to determine the effective responsibility of the camera. Stars are point sources so aperture photometry[23] from astronomical data analysis is adapted to calculate the stars' fluxes. Aperture photometry works as follows. Make a circle around the center star and measure the total data values (or fluxes) of all the inside pixels. The diameter of the circle is larger than the point spread function of the camera. Make a ring outside the circle where no stellar objects are in the ring. Calculate the mean data values within the ring as the background. The difference between the total output and the total background within the circle is the true output value of the center star. The real stellar flux to the true output value is the effective responsibility of the camera. The error of this responsibility is from the signal-to-noise ratio of the star. For example, the previous 250 signal-to-noise ratio of the star will result in an uncertainty of 0.4%. In performing this calculation the dark current of each pixel should be known and subtracted. The nonuniformity of the pixels are also known and corrected. The dark currents and nonuniformity of the camera can be measured in the ground tests. Since the calibration camera is small, the nonuniformity can even be measured onboard with a small flatfield board near the camera. The nonuniformity can also be obtained through earth observation as used by the Hubble space telescope[24]. The nonuniformity is correctable to 0.5% or better. If the dark current of the detectors are much less than the source brightness, which is very likely in visible wavebands and possible in infrareds, the ground measured dark current will not affect the calibration result even without onboard dark current measurement. In general, the calibration errors of the fluxes of ground targets using known flux stars will not be over 1%. The real precision of this process is also dependent on the calibration camera's quality.

Performance degradation is a common issue for the calibration camera working in space. Performance degradation is the change of effective responsibility and nonuniformity. The nonuniformity can be corrected using the methods mentioned in the previous paragraph. The change of the effective responsibility implies that for the same input source flux, the output is different. If the calibration camera's responsibility degrades, the output from the star and that from the ground target will both decrease with same fraction. The ground target's flux in the unit of the stellar flux do not change (refer to equation 7 and explanation there). In fact, even degradation happens to the calibration camera, as long as it is relatively stable within the calibration process, this flux calibration method is valid.

**2.4  The precision of calibrating ground targets with calibrated ground objects**

While the remote sensing camera is performing its normal duty, the calibration camera observes between the stars and the ground targets alternatively. The calibration camera has a small field of view as compared with the remote sensing camera. It can observe a small part of the remote sensing camera's field of view each time. These areas can be directly calibrated by the calibration camera. If the observing areas of the calibration camera are arranged properly, after certain time the remote sensing camera's field of view will be fully covered by the calibration camera so we can use the calibration camera to calibrate the remote sensing camera entirely. This calibration is by nature the vicarious calibration between the calibration camera and the remote sensing camera but it is better than usual vicarious calibration between two satellites. Two satellites usually do not observe the same ground target at the same time while the fluxes from that target changes between the two different times. The observation time and direction between the calibration camera and the remote sensing camera are always same. If the two cameras have same wavebands then the brightness of the observing areas can be calculated directly. If the wavebands between the two cameras are slightly different then the brightness of these areas can be calculated through the cameras' response curves. The calibration precision in this vicarious calibration is mostly determined by the errors of the wavebands and can be better than 0.5%.

The wavebands of the calibration camera and the remote sensing camera may change onboard. The error from this change is determined by the $2^{nd}$ & $3^{rd}$ terms of the right side of equation (10). Since generally most satellites cannot perform onboard wavelength calibrations for the camera, this error is not determinable. This is a problem of this flux calibration. In fact it is a problem for most other satellites too. This problem can only be assured by the manufacture technologies.

If the remote sensing satellite can observe the stars directly, the flux calibration using stars can be performed through observing the stars with remote sensing camera rather than using calibration camera. But the remote sensing camera should have higher sensitivity and good uniformity in doing such observations. High resolution satellite IKONOS used stars for flux calibration[10] and reached a precision of 2.5% difference from that of NASA's calibration method.

**2.5  Total calibration precision using the stars**

The errors and their propagations for remote sensing satellite calibration using stars are discussed in the previous chapters. The flux accuracy of a single standard star is around 1% and the relative flux precisions between different stars are better than 0.2%. The fitting errors to the observational data are around 1-5% while some are better than 1% if the fittings are performed piece by piece. The precision of a group of N stars as a combined calibration source will improve $\sqrt{N}$ times. The calibration precision of the ground target is better than 1% if the target is directly calibrated by the calibration camera. The uncertainty is mostly determined by the manufacture goodness of the calibration camera. The calibration uncertainty of ground target that is recalibrated by the remote sensing camera is less than 0.5%. Put all these errors or uncertainties into equation (10) and calculate the root of squares gives the total error of 1.5-5.1%. Here we omit the contribution from the onboard waveband errors. In general, the flux calibration to the remote sensing satellite with stars can reach a precision of 2-5%. If the best stars are chosen, the precision can reach the level of 2%.

# 3　Conclusion

The flux calibration to the remote sensing satellites with stars is very important to the researches and to the remote sensing applications.1) Stars as natural calibration sources have the best flux accuracies among all the space available natural objects. Their flux accuracy is around 1% and the flux precisions among stars are better than 0.2%. With the stellar flux calibration method we proposed in this paper we can improve the remote sensing calibration accuracy to the levels of 2%. 2) Stars have very stable fluxes and the mean values of hundreds and thousands of stars are even more stable so they can be used as long term calibration sources. This is very important for certain researches such as the long term climate study, the meterological environmental and geological changes due to the sun radiation variabilities. 3) The stars have wide range spectra from ultraviolet to far infrared which cover the main wavebands of the remote sensing satellites so stars can be used to calibrate most remote sensing wavebands. 4) This flux calibration method using stars can greatly reduce the calibration apparatus for the remote sensing satellites and is especially good for large remote sensing cameras. The flux calibration is an independent process and does not take the remote sensing camera's work time. This calibration method can also be a supplement to the normal satellite's calibration method to increase the reliability of the onboard flux calibration.

This project is supported by the Open Project of the State Key Laboratory for Infrared Physics and the Chinese National Science Foundation Grant 11573049. This paper will publish in Chinese on 《Journal of Infrared and Millimeter Waves》 in 2017. The author of this paper translates it into English for the arXiv preprint publication.